\documentclass[fleqn,10pt]{wlscirep}
\usepackage[utf8]{inputenc}
\usepackage[T1]{fontenc}

\usepackage[left]{lineno}
\usepackage{setspace} 
\usepackage[draft]{changes}
\definechangesauthor[color=orange]{Jan}
\definechangesauthor[color=magenta]{Koen}
\definechangesauthor[color=green]{Mat}

\title{The scaling of social interactions across animal species}

\author[a,b,*]{Luis E C Rocha}
\author[b]{Jan Ryckebusch}
\author[a]{Koen Schoors}
\author[c]{Matthew Smith}
\affil[a]{Department of Economics, Ghent University, Ghent, Belgium}
\affil[b]{Department of Physics and Astronomy, Ghent University, Ghent, Belgium}
\affil[c]{The Business School, Edinburgh Napier University, Edinburgh, UK}
\affil[*]{luis.rocha@ugent.be}

\begin{abstract}

Social animals self-organise to create groups to increase protection against predators and productivity. One-to-one interactions are the building blocks of these emergent social structures and may correspond to friendship, grooming, communication, among other social relations. These structures should be robust to failures and provide efficient communication to compensate the costs of forming and maintaining the social contacts but the specific purpose of each social interaction regulates the evolution of the respective social networks. We collate 611 animal social networks and show that the number of social contacts $E$ scales with group size $N$ as a super-linear power-law $E=CN^\beta$ for various species of animals, including humans, other mammals and non-mammals. We identify that the power-law exponent $\beta$ varies according to the social function of the interactions as $\beta = 1+a/4$, with $a \approx {1,2,3,4}$. By fitting a multi-layer model to our data, we observe that the cost to cross social groups also varies according to social function. Relatively low costs are observed for physical contact, grooming and group membership which lead to small groups with high and constant social clustering. Offline friendship has similar patterns while online friendship shows weak social structures. The intermediate case of spatial proximity ($\beta=1.5$ and clustering dependency on network size quantitatively similar to friendship) suggests that proximity interactions may be as relevant for the spread of infectious diseases as for social processes like friendship.
\end{abstract}
\begin{document}

\flushbottom
\maketitle

\thispagestyle{empty}

\section*{Introduction}

Social animals including humans live in groups to optimise the multiplicative benefits of social interactions such as protection, coordination, cooperation, access to information, and fitness, while balancing the competition, disease risk, and stress costs of group living~\cite{Krause2002,molvarCostsBenefitsGroup1994,
Seabright2010}. Social interactions are fundamentally dyadic yet sufficiently diverse to link multiple animals or humans in connected social structures~\cite{Krause2002,Wasserman1994,faustComparingNetworksSpace2002}. The purpose of social interactions is also diverse and spans a range of processes including communication, trust, grooming, dominance, or simply the loosely defined idea of friendship~\cite{Krause2002,Wasserman1994,Caroline1993}. Correlations between social interactions, as for example dominance and physical contact, friendship ties maintained through communication, or the intertwined relation between trust and spatial proximity, reveal the complexity of social phenomena and suggest that common principles may underlie the formation of social ties. 

A fundamental question concerns how the number of social connections depends on group size, and whether there are any emerging patterns in this relationship. The answer may reveal whether interaction patterns become more complex with size in order to maintain efficient social structures within the group. The cost to establish and maintain social contacts in small groups is relatively low but increases in larger groups~\cite{sueur2011group}. This increasing costs leads to peer selection, either by necessity or affinity, up to a species-specific cognitive saturation point in the number of contacts one can manage~\cite{Dunbar1992}. Assuming that all members of a social group are reachable via social ties, in the limiting scenarios, a group of size $N$ individuals may have a fragile star-like structure with $E=N-1$ social ties to minimise social interactions (lowest cost) or a fully connected clique with $E=N(N-1)/2$ ties (highest cost). 

Evolutionary arguments support that social groups specialise and optimise social interactions to save resources while keeping or increasing the group efficiency~\cite{terborgh_socioecology_1986,guindre-parker_survival_2020,pasquaretta2014social}, as for example in response to predators (ecological conditions)~\cite{williams_optimal_2003} or to fitness~\cite{Roberts2020, Dunbar2018}. There is also the argument that human social networks have an optimal size to optimise information transfer within groups~\cite{West2020}. Research on urban systems shows that human societies also organise in groups (e.g.\ cities) to optimise resources like infra-structure and to increase intellectual, social and economic outputs~\cite{Bettencourt2007, Bettencourt2013}. These observations lead us to hypothesise that across species and social contexts, the number of social contacts $E$ scales with group size $N$ as $E = C N^{\beta}$, where $C$ and $\beta$ are positive constants.

Until recently, measuring social interactions was laborious. Past research relied on observations of animal and human behaviour or self-reporting of social contacts through questionnaires~\cite{Wasserman1994}. A natural limitation of these techniques is the size of the observed populations and potential recalling errors, as for example the inability to accurately identify or quantify each interaction~\cite{Krause2002, Wasserman1994}. Electronic devices (e.g.\ mobile phones~\cite{Eagle2004} or proximity sensors~\cite{Barrat2014,migliano2017characterization}) and online platforms now provide means for passive and accurate recording of spatio-temporal location, communication between animals and between humans, among other forms of animal or human interactions. State-of-the-art electronic data collection is scalable but its ability to detect authentic social interactions may be questioned and should be treated cautiously~\cite{Kibanov2015, Lieberman2020}.

We collate extensive data to show empirically that the number of social contacts scales super-linearly (i.e.\ $\beta>1$) with group size and that social interactions can be categorised in different exponents $\beta$ independently of the animal species. We provide evidence that this scaling is necessary to maintain fundamental complex network structures irrespective of existing group sizes. We also fit our data to a social network model and show that a multi-layer structure and the cost of crossing social layers may explain the estimated scaling exponents.

\begin{table*}[htb]
\begin{center}
\caption{Number of networks for each type of social interaction and animal class. In a total of 611 networks, there are 179 cases of human and 432 cases of non-human social interactions, including 281 captive and 151 free-ranging animals.}
\begin{tabular}{l|ccccccc|c}
\hline
  Social interaction type     & \shortstack{Mammalian \\ non-primates} & \shortstack{Mammalian \\ primates} & \shortstack{Mammalian \\ Humans} & Actinopterygii & Aves & Insecta & Reptilia & Total \\
\hline
A: physical contact   &  0 &  4 &   0 &  0 & 2 & 244 & 0 & 250 \\
B: grooming           &  0 & 23 &   0 &  0 & 0 &   0 & 0 &  23 \\
C: group membership   &  4 &  0 &   0 &  7 & 5 &   0 & 0 &  16 \\
D: spatial proximity  & 63 & 58 &  88 &  9 & 0 &  12 & 1 & 231 \\
E: offline friendship &  0 &  0 &  67 &  0 & 0 &   0 & 0 &  67 \\
F: online friendship  &  0 &  0 &  24 &  0 & 0 &   0 & 0 &  24 \\ \hline
Total              & 67 & 85 & 179 & 16 & 7 & 256 & 1 & 611 \\
\hline
\end{tabular}
\label{tab:01}
\end{center}
\end{table*}

\section*{Results}

\begin{figure*}[thb]
\centering
\includegraphics[scale=0.6]{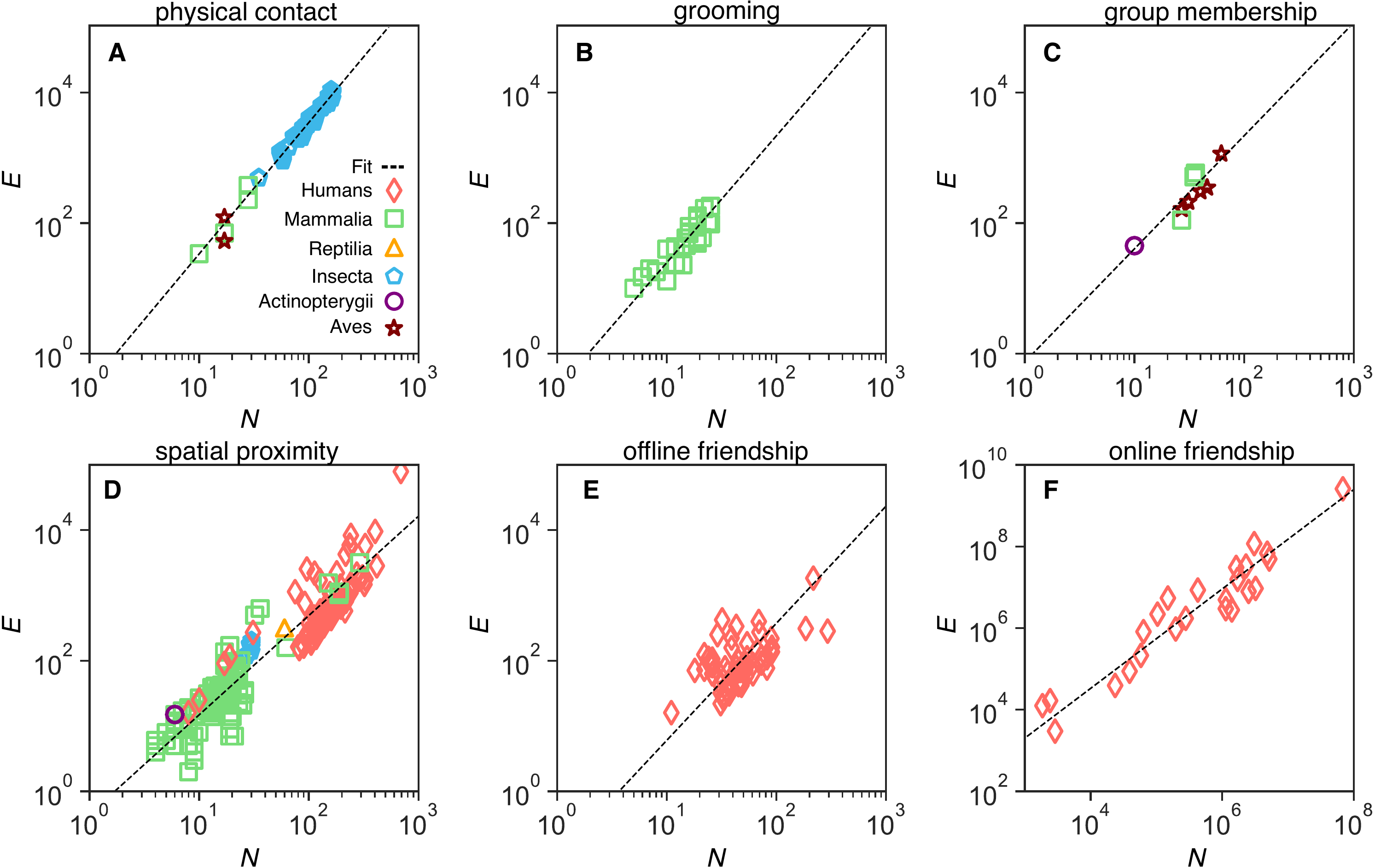}
\caption{The number of social connections $E$ versus the group size $N$ across species. Empirical data (each symbol corresponds to a different species) and regression curves (dashed lines) for all 6 categories of social interactions: (A) physical contact ($\hat{\beta}_{\text{A}} = 2.01$, $95\%$CI $[1.98,2.04]$, $n=250$); (B) grooming ($\hat{\beta}_{\text{B}} = 1.94$, $95\%$CI $[1.53,2.36]$, $n=23$); (C) group membership ($\hat{\beta}_{\text{C}} = 1.73$, $95\%$CI $[1.47,1.99]$, $n=16$); (D) spatial proximity ($\hat{\beta}_{\text{D}} = 1.52$, $95\%$CI $[1.45,1.59]$, $n=231$), with $38\%$ human and $62\%$ non-human networks; (E) offline friendship ($\hat{\beta}_{\text{E}} = 1.79$, $95\%$CI $[1.30,2.29]$, $n=67$); (F) online friendship ($\hat{\beta}_{\text{F}} = 1.22$, $95\%$CI $[1.06,1.37]$, $n=24$). Details of the fitting in Table~\ref{tab:02}. All axes are in log-scale. }
\label{fig:01}
\end{figure*}

The data sets were collated using online databases of animal and human social networks previously analysed by other authors. All networks were reviewed for consistency and the data sets were standardised such that only unique pairs of social contacts were counted, i.e.\ self-loops, weighting, timings of contacts, and directions were removed. Social interactions were identified and labelled in the original studies by domain experts via direct observation (animal interactions), questionnaires (offline friendship), electronic devices (spatial proximity), and online platforms (online friendship) (see SI). To minimise potential ambiguities, each network was constructed based on the specific definition of social interaction in the respective original study. Table~\ref{tab:01} shows the number of networks for each type of social interaction and animal class, including captive and free-ranging animals. The network size varies across species and social interactions because of experimental settings, characteristics and limitations of the study populations, e.g.\ the observation capacity of researchers, cost of technical devices, free-range vs. confined animals, online platforms, or animals living in small groups (see SI).

\subsection*{Scaling of social interactions}

The networks of social interactions were grouped in categories following the type of social interactions as reported in the original studies (Table~\ref{tab:01}). Figure~\ref{fig:01} shows the scaling between the number of social contacts $E$ and size $N$ (i.e.\ the number of interacting individuals) for each of the 6 original categories. We assume that the scaling of social relations is independent of species and test our hypothesis $E=CN^\beta$ by fitting a power-law to the data using logarithmic transformed variables to evenly distribute the data points:
\begin{equation}
      \log E = \log C + \beta \log N \; .
\end{equation}
The fitting exercise gives super-linear power-law exponents (i.e.\ $\beta>1$) and strong linear correlations ($0.55<r<0.99$) for all categories of social interactions (Table~\ref{tab:02}). Assuming a small error $\epsilon$ in $\hat{\beta}$, the exponents follow the general law ($\beta=1+a/4$) with $a \approx 1$ for online friendship ($\epsilon=12\%$), $a \approx 2$ for spatial proximity ($\epsilon=4\%$), $a \approx 3$ for group membership ($\epsilon=2.6\%$) and offline friendship ($\epsilon=5\%$), and $a \approx 4$ for physical contacts ($\epsilon=1\%$) and grooming ($\epsilon=6\%$), despite differences in species and sample sizes (Fig.~\ref{fig:01}).

\begin{table*}[htb]
\begin{center}
\caption{Best fitting exponents for the 6 types of social interactions. The variable $n$ gives for each type of social interaction the number of different networks that was included in the fit. Orthogonal regression is used to account for measurement errors in both axes. }
\begin{tabular}{lccccccc}
\hline
 Social interaction type & $n$ & \multicolumn{2}{c}{$\hat{\beta}$} & \multicolumn{2}{c}{$\hat{C}$} & Pearson     & $p$-value \\
 & & best fit      & $95\%$CI & best fit & $95\%$CI & correlation & \\
\hline
A: physical contact   & 250 & 2.01 & [1.98,2.04] & 0.75 & [0.33,1.68] & 0.96 & <.01 \\
B: grooming           & 23  & 1.94 & [1.53,2.36] & 0.28 & [0.09,0.86] & 0.82 & <.01 \\
C: group membership   & 16  & 1.73 & [1.47,1.99] & 0.33 & [0.29,0.37] & 0.89 & <.01 \\
D: spatial proximity  & 231 & 1.52 & [1.45,1.59] & 0.44 & [0.34,0.58] & 0.55 & <.01 \\
E: offline friendship & 67  & 1.79 & [1.30,2.29] & 0.10 & [0.01,0.65] & 0.57 & <.01 \\
F: online friendship  & 24  & 1.22 & [1.06,1.37] & 0.44 & [0.06,3.36] & 0.99 & <.01 \\ \hline
\hline
\end{tabular}
\label{tab:02}
\end{center}
\end{table*}

This super-linear scaling indicates increasing densification of social contacts, that is, larger social groups have on average more social contacts per-capita than the smaller ones. It is not surprising that $\beta>1$ because the number of social connections must scale at least linearly with group size ($E \propto N$) to maintain the social network connected; this is known as the percolation threshold in random networks~\cite{Newman2010}. If $E \approx N$, small perturbations may fragment the network, breaking down the group structure. Furthermore, $\beta>1$ suggests that a super-linear number of contacts are necessary to create and maintain the complex social network structures for the groups to function cost-efficiently irrespective of size.

\subsection*{Social network structure}

\begin{figure*}[!thb]
%\centering
\includegraphics[scale=0.56]{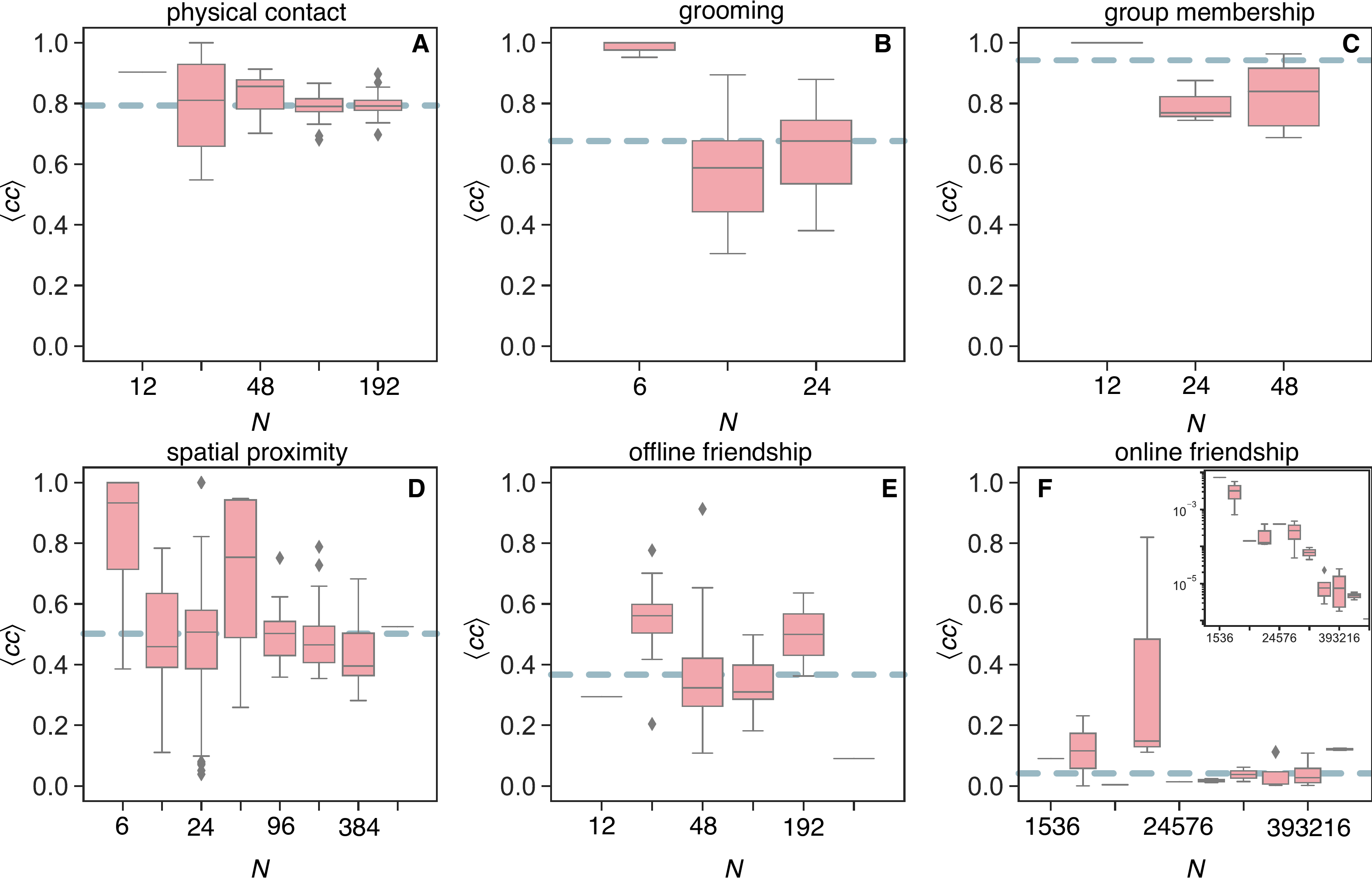}
\caption{Network clustering structures. The average clustering coefficient $\langle cc \rangle$ between close contacts vs. network size for (A) physical contact (Median values for the empirical $M_{\text{emp}}=0.94$ and randomised $M_{\text{rand}}=0.88$ versions of the same networks); (B) grooming ($M_{\text{emp}}=0.68$ and $M_{\text{rand}}=0.56$); (C) group membership ($M_{\text{emp}}=0.79$ and $M_{\text{rand}}=0.69$); (D) spatial proximity ($M_{\text{emp}}=0.50$ and $M_{\text{rand}}=0.22$); (E) offline friendship ($M_{\text{emp}}=0.37$ and $M_{\text{rand}}=0.07$); (F) online friendship ($M_{\text{emp}}=0.04$ and $M_{\text{rand}}=3.5 \cdot 10^{-5}$); the inset is the distribution for the random version of the same networks. Dashed horizontal lines are the median values of the empirical networks. Log-binned (x-axes) Tukey box plots with diamonds representing outliers. }
\label{fig:02}
\end{figure*}

We study the network structures for each of the six types of social interactions (See Methods). The clustering coefficient $\langle cc \rangle$ is a local measure of the level of sociality between common contacts of a focal individual (i.e.\ the fraction of social triangles). Its intensity indicates an evolutionary group advantage as for example fitness benefits~\cite{deliusTransitiveRespondingAnimals1998,cheneyNetworkConnectionsDyadic}. Networks with higher clustering are relatively more robust since the deletion of a social connection would not significantly affect interaction and communication among close contacts. In our social networks, $\langle cc \rangle$ is constant for varying network size for all types of social contacts (Fig.~\ref{fig:02}). In random networks, the clustering coefficient decays with increasing network size as $\langle cc \rangle = \langle k \rangle /N$, where $\langle k \rangle$ is the average number of contacts (or edges) in the network~\cite{Newman2010}. The inset of Fig.~\ref{fig:02}F shows the results for the randomised versions of the same online friendship networks (see SI for the other categories). In all categories, there is a higher clustering coefficient than expected on the basis of randomised social contacts (See caption Fig.~\ref{fig:02}). Since the average degree is defined as $\langle k \rangle = 2E/N$, we have $\langle cc \rangle = \langle k \rangle /N= 2E/N^2$ and thus would need $E \propto N^2$ to have constant clustering in random networks. Evolutionary theory implies that more complex structures may emerge in such social systems to optimise resources, e.g.\ to reap the fitness related benefits, and thus relatively less social contacts become necessary to reach the same level of clustering across group sizes~\cite{deliusTransitiveRespondingAnimals1998, cheneyNetworkConnectionsDyadic}. For example, for some classes of random heterogeneous networks, $\langle cc_{\text{h}} \rangle = A/N$, where the proportionality constant $A$ depends on the heterogeneity of the distribution of contacts among individuals and is lower than $\langle k \rangle$~\cite{Newman2010}.

\begin{figure*}[!thb]
%\centering
\includegraphics[scale=0.56]{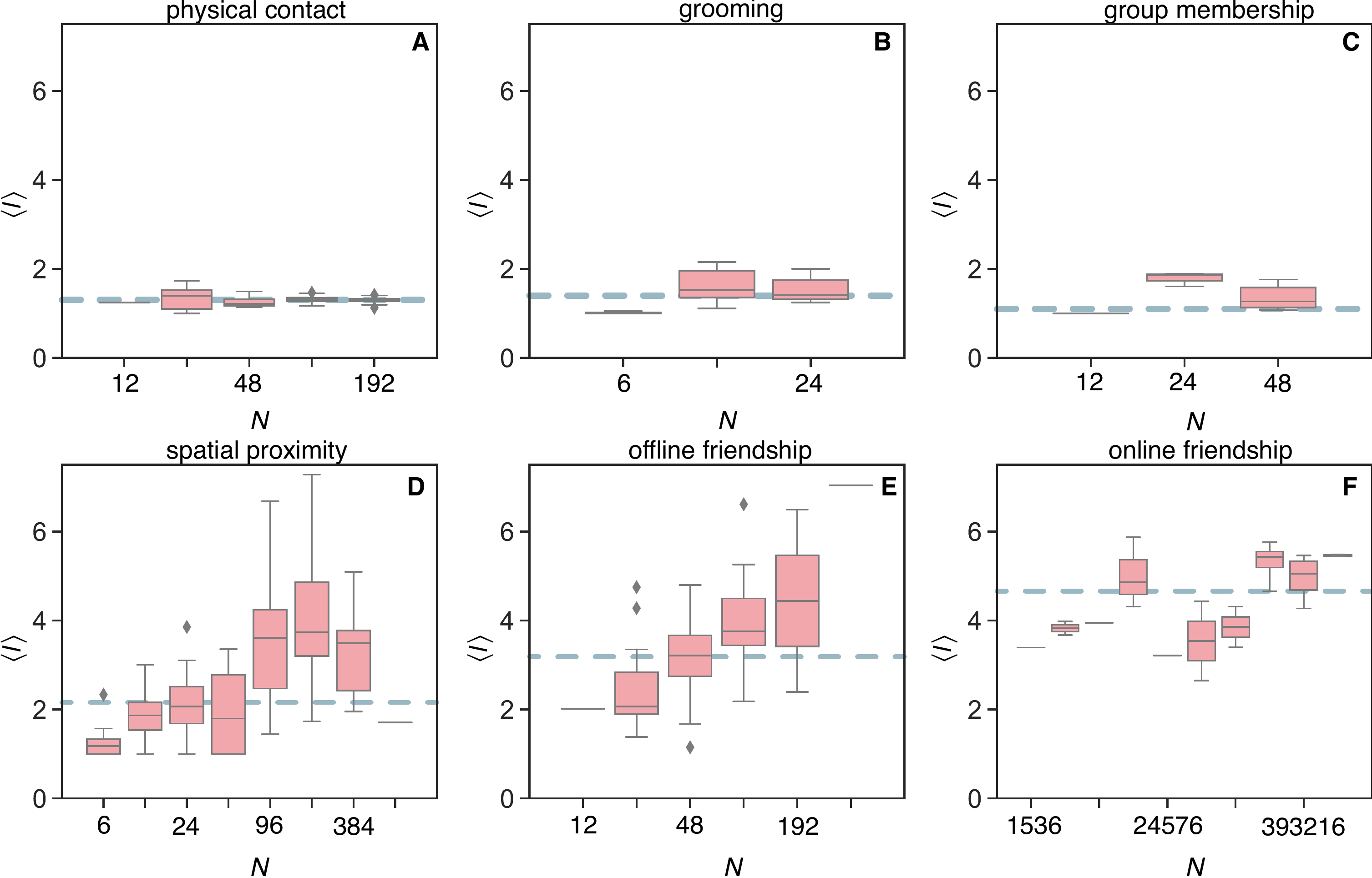}
\caption{Network path structures. The average shortest path-length $\langle l \rangle$ vs. network size in the networks of (A) physical contact (Median values for the empirical $M_{\text{emp}}=1.10$ and randomised $M_{\text{rand}}=1.05$ versions of the original network); (B) grooming ($M_{\text{emp}}=1.39$ and $M_{\text{rand}}=1.29$); (C) group membership ($M_{\text{emp}}=1.30$ and $M_{\text{rand}}=1.09$); (D) spatial proximity ($M_{\text{emp}}=2.16$ and $M_{\text{rand}}=2.00$); (E) offline friendship ($M_{\text{emp}}=3.19$ and $M_{\text{rand}}=3.36$); (F) online friendship ($M_{\text{emp}}=4.66$ and $M_{\text{rand}}=4.97$). Dashed horizontal lines are the median values of the empirical networks. Log-binned (x-axes) Tukey box plots with diamonds representing outliers. }
\label{fig:03}
\end{figure*}

The average length of the shortest-paths $\langle l \rangle$ measures the average distance between any pairs of individuals in the social network and quantifies the communication potential between parts of the network~\cite{Latora2001}. Shorter average distances (i.e.\ $\langle l \rangle \ll N$, resulting in the small-world effect~\cite{Travers1969}) indicate that information flows quickly over the network, which is a fundamental characteristic of efficient group organisation~\cite{schnettler2009structured}. For physical contacts, grooming, and group membership, $\langle l \rangle$ is constant and slightly higher than one (Fig.~\ref{fig:02}A-C). For spatial proximity and offline friendship, the values increase with size following quantitatively similar trends (Fig.~\ref{fig:02}D-E). The results for online friendship suggest a constant trend (Fig.~\ref{fig:02}F). In all cases, the average path-length is $\langle l \rangle < 6$, which is the small-world horizon observed empirically~\cite{Newman2010}. For all 6 categories, the random versions of the same networks give constant relations albeit generally with lower values (see SI). In theoretical random networks, the average distance increases slowly with the network size as $\langle l \rangle \approx \log(N) / \log(\langle k \rangle)$~\cite{Newman2010}. Nevertheless, the average path-length $\langle l \rangle$ is approximately constant across group sizes if $E \propto N^{\beta}$ for $\beta>1$, since in this case $\langle l \rangle \approx \log(N) / \log (2N^{\beta-1}) \approx 1/(\beta-1)$. Smaller $\beta$ thus leads to higher $\langle l \rangle$, as observed in the analysed networks. In some classes of heterogeneous random networks, $\langle l \rangle$ is also nearly constant with network size~\cite{Newman2010}. The density of contacts explains the low $\langle l \rangle$ for physical contact, grooming and group membership. The discrepancy of spatial proximity, offline and online friendship with the random case indicates that more complex network structures are being formed in larger groups for these types of social interactions. In sparse networks, like those, a high level of local clustering increases the distance between random pairs of network nodes because of local spots of connectivity redundancy~\cite{Newman2010}. Taken together, the constant clustering across network sizes (Fig.~\ref{fig:02}) implies that the average distance will necessarily increase (Fig.~\ref{fig:03}), unless followed by a sufficient increase in the number of connections (to maintain low average distances as the group increases). The growth in offline friendship followed by a seemingly constant pattern for online friendship (which has larger sizes) suggests a potential saturation in $\langle l \rangle$ for human friendship in line with the small-world horizon observed in previous studies~\cite{Newman2010}. Although communication remains efficient (because $\langle l \rangle \ll N$), the benefits of forming larger groups do not compensate the costs of optimising certain network structures, as is the case for other types of social interactions involving physical contact.

\subsection*{Multi-layer model}

Multi-layer models can be used to represent the underlying generative mechanisms through which individuals combine skills and affinity to build up more complex social groups. From single individuals to the entire population, individuals may be stratified in layers (or levels) corresponding to different groups~\cite{Watts2002}. For example, living in households (layer 1) within neighbourhoods (layer 2) that in turn are part of cities (layer 3), and so on, seems natural for humans. While people mostly interact with those in the same group (e.g.\ within the same household), interactions across groups are less frequent~\cite{Dunbar2020} (e.g.\ between different households in the same neighbourhood). Interactions across groups at the same layer are necessary to define higher-order groups, i.e.\ a group at the next higher layer, as for example a neighbourhood is a result of interactions between individuals from different households. Multi-layer models have been used to explain spatial relations in vascular~\cite{West1999} and infrastructure~\cite{Bettencourt2013} systems. We argue that such models are also of value for social groups, not necessarily spatially bound, since multi-layer organisation has been observed across animal species in which a relation between group sizes in different layers vary from nearly 2.5 in primates to about 3 for other mammals including humans~\cite{Hill2008,Dunbar2018}. This means that individuals are organised as multiples of 3, for example, in groups of 5 (layer 1), 15 (layer 2), 45 (layer 3), and so on. The model detailed below does not aim to reproduce all structures of the 611 analysed networks but focuses on the scaling exponents $\beta$.

\begin{figure}[!thb]
%\centering
\includegraphics[scale=0.83]{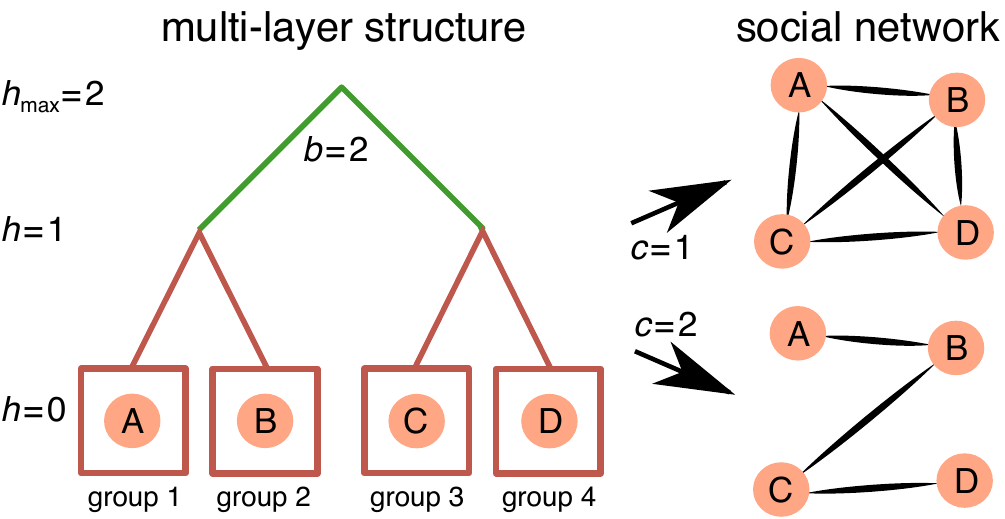}
\caption{Multi-layer social model. The underlying multi-layer structure (left) defines the probability $p_{\Delta h}(i,j)\propto c^{-\Delta h}$ of forming connections between individuals $i$ and $j$ in the social network (right). If $c=1$, everyone interacts with everyone else leading to a fully connected network whereas for higher $c$, interactions between closer individuals (lower $h$) are more common. For example, the distance $\Delta h_{A,B}=1$ and $\Delta h_{A,D}=2$. With cost $c=1$, edges $(A,B)$ and $(A,D)$ are equally likely ($p_1 = p_2 \propto 1$), whereas with higher cost, e.g.\ $c=2$, edge $(A,B)$ is more likely ($p_1 \propto 1/2$) to occur than edge $(A,D)$ ($p_2 \propto 1/4$).}
\label{fig:04}
\end{figure}

The self-similar multi-layer group structure is mathematically represented as a branching tree with a group at layer $h$ split into $b$ sub-groups at layer $h-1$ (Fig.~\ref{fig:04}). At the highest layer $h_{\text{max}}$, all individuals belong to a single social group, i.e.\ $N=b^{h_{\text{max}}}$, and at the lowest layer ($h_{\text{min}} = 0$), each group is formed by a single unique individual. In this model, individuals $i$ and $j$ make a social contact $(i,j)$ with probability $p_{\Delta h}(i,j)$ dependent on the distance $\Delta h$ between the layers that separate them. Closer individuals (e.g.\ at distance $\Delta h=1$ because they are living in the same neighbourhood or belonging to the same social group) are more likely to interact than individuals living far apart (e.g.\ at distance $\Delta h=2$ because they are living in different cities or belonging to different social groups), i.e.\ $p_{\Delta h}(i,j)$ decreases with $\Delta h$. The multi-layer tree-like structure only defines the distance $\Delta h$ between the groups that is in turn used to form contacts in the social network (Fig.~\ref{fig:04}); the resulting social network only has tree-like structure for sparse networks, i.e.\ when $E \ll N^2$. The self-similarity between layers implies that $p_{\Delta h}(i,j)/p_{\Delta h-1}(i,j)=const$~\cite{Leskovec2005}. A power-law of the form $p_{\Delta h} \propto c^{-\Delta h}$, with $c>1$, satisfies this relationship. The parameter $c$ represents the cost to make social interactions across layers, that we assume is lower than the cost to create a new layer, i.e.\ $c < b$, because multiple contacts are necessary to establish a new layer. For a given individual $i$, the expected number of social connections $\langle e \rangle_i$ is:
\begin{align*}
\langle e \rangle_i &= \sum_{j \neq i} p_{\Delta h}(i,j) = \sum_{\Delta h=1}^{h_{\text{max}}} (b-1)b^{\Delta h-1}c^{-\Delta h} = \frac{b-1}{c} \sum_{\Delta h=1}^{h_{\text{max}}} (b/c)^{\Delta h-1}.
\end{align*}
For $1 \leq c < b$, the sum converges:
\begin{align*}
%\langle E \rangle =& \dfrac{b-1}{c} \dfrac{(b/c)^{\log_b (N)}-1}{ (b/c) -1} = \dfrac{b-1}{b-c} \left( N^{1-\log_b (c)}-1 \right) \\
\langle e \rangle_i = \frac{b-1}{c} \times \frac{(b/c)^{h_{\text{max}}}-1}{(b/c)-1}
\end{align*}
Since $N = b^{h_{\text{max}}}$ and $c^{h_{\text{max}}} = b^{h_{\text{max}} \log_b (c)}$, we get:
\begin{align*}
\langle e \rangle_i &\propto N^{1-\log_b (c)}.
\end{align*}
Therefore, the total number of social connections is:
\begin{align*}
E \sim N \times \langle e \rangle_i \propto N \times N^{1-\log_b (c)} \sim  N^{2-\log_b (c)}.
\end{align*}
The multi-layer model implies that $\beta=2-\log_b (c)$. Assuming that $b=2.5$~\cite{Dunbar2018}, the cost of connections is thus $c_{\text{A}} = 1$ for physical contact ($\beta_{\text{A}} \approx 2$), $c_{\text{B}} = 1$ for grooming ($\beta_{\text{B}} \approx 2$), $c_{\text{C}} = 1.26$ for group membership ($\beta_{\text{C}} \approx 1.75$), $c_{\text{D}} = 1.58$ for spatial proximity ($\beta_{\text{D}} \approx 1.5$), $c_{\text{E}} = 1.26$ for offline friendship ($\beta_{\text{E}} \approx 1.75$) and $c_{\text{F}} = 1.99$ for online friendship ($\beta_{\text{F}} \approx 1.25$). This cost is associated to crossing (virtual) barriers between social groups that might cause the creation of larger groups. The low cost ($c = 1$) for physical contact and grooming means that $p_{\Delta h} = 1$, i.e.\ the probability to form connections is independent of the social distance $\Delta h$, collapsing the assumption of multi-layer structure. For such types of social interactions, the connections within the same social group are favoured; individuals do not groom in different social groups nor make persistent physical contacts, except physical contacts for conflict that would not be reflected in our data. This effect is related to the high clustering coefficient reported in previous sections and may explain the relatively small size of such networks. The number of social contacts for such activities is limited within the same social group. 

Online friendship, on the other hand, is costly ($c \approx 1.99$) in terms of crossing social boundaries to connect individuals from different social groups~\cite{Tamarit2018}, e.g.\ with different tastes, ideas, location, age, and so on. Socially closer individuals would be favoured here as well since it is harder to be friends with dissimilar people than with those similar to each other~\cite{Dunbar2020}. However, given that online connections are cheap to establish and maintain (i.e.\ do not need nurturing and resources), the multi-layer structure becomes relevant with a non-negligible number of socially distant connections being formed. Furthermore, online friendship typically mixes (real) friends, acquaintances, relatives, and co-workers, each belonging to different social groups, with some individuals acting as social brokers. For example, online friendship is more easily established between those studying in the same school than at different schools; however, inter-school friendship is facilitated by the online platform, though socially costly (lack of face-to-face interactions, no common friends, building trust). For networks derived from mobile phone communication in urban populations, a scaling exponent $\beta=1.15$ has been reported~\cite{Calabrese2011, Pan2013, Schlapfer2014}. Such mobile communication data sets mix professional and personal relations which possibly also leads to higher costs in the sense of crossing social boundaries. In one study, a constant clustering coefficient has been also observed suggesting that similar underlying principles may explain the formation of such social or communication structures~\cite{Schlapfer2014}. The multi-layer structure becomes less relevant for offline friendship ($c \approx 1.32$) that are typically more spatially constrained in our data. For example, students or prison inmates will report friendship with those around them. In schools, from where most of our data come from, the social structure is seen at the class and school layers only. Given experimental limitations, it is often not possible to report friends outside the study setting, which could reveal higher social layers, e.g.\ neighbourhood friends. It is possible that the exponent $\beta$ for friendship is thus between what we estimated for offline and online friendship if all layers of friends and not only those in the same study setting were reported.

Our analysis finds an intermediate exponent ($\beta = 1.5$) and cost ($c \approx 1.58$) for spatial proximity. Spatial proximity is a particular type of social interaction. Grooming, physical contact and human friendship are well-defined interactions identified, respectively, by observing joint activities or by directly inquiring individuals. However, spatial proximity interactions are measured by sensors or direct observation and capture a mixture of social situations. Spatial proximity might reflect affinity, trust and friendship between individuals and animals sharing the same space~\cite{Dunbar2020}, e.g.\ persistent spatial proximity between pairs of cows~\cite{Rocha2020}, or behavioural or trait similarity, i.e.\ homophily, as for example friends visiting a museum~\cite{Isella2011} or health-care workers in hospitals~\cite{Vanhems2013}. On the other hand, spatial proximity interactions might simply reflect spatial constrains forcing individuals and animals to be in close proximity during periods of time, e.g.\ a group of visitors of an art exhibition~\cite{Isella2011} or confined animals~\cite{Rocha2020}. Nevertheless, also in the later, affinity and trust are reflected in the proximity contacts. As discussed above, it is possible that friendship at the society layer likely follows patterns intermediate to those observed in the online ($\beta=1.25$) and offline ($\beta=1.75$) categories. The existing literature associating friendship to time that individuals spent together~\cite{Dunbar2020} and the observation that spatial proximity contacts follow an intermediate exponent ($\beta=1.5$) suggest a potential link between these social interactions. We cannot make a strong association between the two types of social interactions due to lack of data of offline friendship in larger populations. Previous modelling exercises in urban populations suggest that $\beta_{B} = 1.5$ can be explained by mobility ($H=2$, where $H$ is the Hausdorff dimension of a path in space) over two dimensional ($D=2$) spaces based on the assumption that fully-mixed populations may fully explore a given area~\cite{Bettencourt2013}. While this assumption may hold within, e.g.\ schools, museums or barns, it does not apply on larger spatial areas since humans and animals are territorial and tend to spend most of time within certain locations~\cite{Song2010} or with certain individuals~\cite{Dunbar2020}. On the other hand, the same model suggests that contacts per-capita scale as $0.25$ (i.e.\ $\beta = 1.25$) under the same conditions (i.e.\ $H=D=2$). This fits well to our findings for offline friendship, where people may virtually explore the whole social space and potentially interact with different individuals.

\section*{Conclusions}

Our findings reveal key aspects of the organisation of animal social networks. Though primates and non-primates (including humans) are more represented than other animals in our data set, the universal scaling relations $E=CN^{\beta}$ between the number of social contacts $E$ and size $N$ suggest common organisation principles across animal species that can be explained by multi-layer models designed to maintain the functioning of the social groups~\cite{Hill2008, Dunbar2018}. Different scaling exponents following the general relation $\beta = 1+a/4$, with $a \approx {1,2,3,4}$ allow us to distinguish types of social interactions and to infer network structures underlying those interactions. For all types of social interactions, the local clustering remains constant for increasing network sizes albeit having different intensity in each case. Physical contacts, grooming and group membership have similar constant median values that are higher than observed for spatial proximity, offline and online friendships. The average path-length is also constant and follow the small-world pattern (i.e.\ $\langle l \rangle \ll N$) for most cases with the exception of spatial proximity and offline friendship where a quantitatively similar positive trend is observed with values below the small-world horizon of $\langle l \rangle \approx 6$ previously observed in social networks~\cite{Newman2010}.

One may argue that humans differ from other animals by developing more efficient social network structures, with relatively less contacts for higher network sizes, and thus lowering the scaling exponents. There is a quantifiable relationship with brain and group sizes, along with the complexity of the interactions. Humans are able to process the cognitive demand of other forms of relationships such as friendship, rather than mating and dominance relations that often occur within other animals and species~\cite{dunbar_social_2009}. The common scaling pattern observed across species and particularly for spatial proximity weakens the hypothesis that animals differ. Our results suggest that the type of social interaction, and to a lesser extent, the group size, are more relevant to determine the scaling exponents than the animal species. We reached this conclusion by combining data from different species. More statistical power could be achieved with a larger sample of network data for specific combinations of social interactions and species in order to study these relations separately. Given the multi-layer structure of social networks and experimental constraints, offline friendship data sets are limited to relatively small social circles~\cite{Dunbar2020}. If one could map higher social layers, the scaling exponent could decrease, likely to the same value as observed for spatial proximity. If this is confirmed in future studies, we will be able to infer that spatial proximity is a proxy of friendship across animals species~\cite{Dunbar2020}.

Physical contact, grooming and group membership are associated with more robust and topologically efficient networks (since clustering is higher and path-lengths are shorter) than friendship and proximity interactions. This social cohesion is a result of homophily and coordination to maintain group functioning, which likely creates smaller groups in these categories relative to friendship and proximity categories because of the cost of nurturing contacts. The frequency and number of social interactions leading to stable social contacts are also important to regulate diffusion processes such as communication~\cite{Latora2001}, innovation~\cite{Bettencourt2007,Bettencourt2013}, infectious diseases~\cite{Barrat2014, Rocha2015} and social phenomena~\cite{Dunbar2020,Alves2013}. Our results suggest that physical contacts and grooming are more efficient than proximity to facilitate spread phenomena at the population (network) level. Online friendships are associated to looser social structures easier to fragment as the groups increase in size. The relatively high cost of nurturing too many online social contacts across social layers restrains the opportunities to generate higher clustering or common friends, and create redundant structures, as observed in the smaller networks related to activities necessary to keep the group functioning.

Although we focus on temporally stable social networks~\cite{Rocha2017}, the availability of temporal information and intensity of certain social interactions could also help to understand the formation and dissolution of social contacts and how particular network structures are formed. Future research should add a quality measure to social interactions (e.g.\ via weights or temporal dynamics) to investigate the varying importance of creating and maintaining particular structures~\cite{Farine2015}. Strong super-linear scaling implies prohibitive social costs to maintain larger groups for some types of social interactions. The questions on whether there is a maximum or optimal group size in which efficient groups can exist and fitness is maximised~\cite{ward_evolution_2016}, or whether more complex network structures are necessary to sustain larger groups, remains open.

\section*{Methods}

\subsection*{Data}

The data sets used in this study were collected using public network data repositories. A list of repositories and a full list of the original references for the 611 data sets are available in the SI. The 6 types of social interactions: physical contact, grooming, group membership, spatial proximity, offline friendship and online friendship were identified and labelled in the original studies by domain experts via direct observation (animal interactions), questionnaires (offline friendship), electronic devices (spatial proximity), and online platforms (online friendship). All 611 networks were standardised for the analysis, including the removal of self-loops, edge directions, and edge weights.

\subsection*{Networks}

A network $G$ of size $N$ is defined as a set of $N$ nodes $i$ and a set of $E$ edges $(i,j)$ connecting nodes $i$ and $j$. A node represents either a person or an animal. An edge represents a social connection of a specific type. In an undirected network, edges are reciprocal, i.e.\ $(i,j)=(j,i)$. In a network without self-loops, there is no edge $(i,i)$.

The clustering coefficient of a node $i$ is given by:
\begin{equation}
    cc_i = 2e_i/(n_i (n_i-1)) \; ,
\end{equation}
where $e_i$ is the number of edges between the $n_i$ nodes directly connected to node $i$. The average clustering coefficient of the network $G$ is thus:
\begin{equation}
    \langle cc \rangle = \frac{1}{N} \sum_{i=1}^N cc_i \; .
\end{equation}
The topological distance between the nodes $i$ and $j$ is the length of the shortest-path $l_{ij}$ in number of edges. It is calculated within the largest connected component of the network $G$. In the largest connected component, there is at least one path between any pairs of nodes $i$ and $j$. The average shortest-path length is:
\begin{equation}
    \langle l \rangle = \frac{1}{N(N-1)} \sum_{i,j=1}^{N} l_{ij} \; .
\end{equation}

\bibliography{scaling}

\section*{Acknowledgements}

The authors thank Luana de Freitas Nascimento for helpful discussions.

\section*{Author contributions statement}

L.R. designed the research, made the analysis and wrote the draft; J.R., K.S., M.S. contributed with methods; All authors revised the manuscript.

\section*{Additional information}

There are no competing interests.

\end{document}